\documentstyle [a4,11pt]{article}
\def\spose#1{\hbox to 0pt{#1\hss}}
\def\simlt{\mathrel{\spose{\lower 3pt\hbox{$\mathchar"218$}}
     \raise 2.0pt\hbox{$\mathchar"13C$}}}
\def\simgt{\mathrel{\spose{\lower 3pt\hbox{$\mathchar"218$}}
     \raise 2.0pt\hbox{$\mathchar"13E$}}}
\def\lsim{\rlap{$<$}{\lower 1.0ex\hbox{$\sim$}}}
\def\gsim{\rlap{$>$}{\lower 1.0ex\hbox{$\sim$}}}
\newcommand{\etal}{{et al.}~}
\newcommand{\done}{\delta^{(1)}}
\newcommand{\de}{\delta}
\newcommand{\p}{\partial}
\newcommand{\tri}{\triangle}
\newcommand{\f}{\frac}

\newcommand{\Om}{\Omega}
\newcommand{\s}{\sigma}
\newcommand{\al}{\alpha}
\newcommand{\fd}{\tilde{\delta}}
\newcommand{\fv}{\tilde{v}}

\newcommand{\bfx}{{\bf x}}

\newcommand{\bfk}{{\bf k}}
\newcommand{\bfv}{{\bf v}}
\newcommand{\bfq}{{\bf q}}
\newcommand{\bfg}{{\bf g}}
\newcommand{\bfF}{{\bf F}}
\newcommand{\bc}{\begin{center}}
\newcommand{\be}{\begin{equation}}
\newcommand{\ee}{\end{equation}}
\newcommand{\ec}{\end{center}}
\newcommand{\lan}{\langle}
\newcommand{\ran}{\rangle}

\title{{\LARGE {\bf KURTOSIS AS A NON--GAUSSIAN SIGNATURE OF
THE LARGE--SCALE VELOCITY FIELD}}\\}
\vspace{1cm}
\author{ {\bf Paolo CATELAN}$^{\,1,2}$ \&
{\bf Lauro MOSCARDINI}$^{\,3}$ \\ \\
$^{1\,}$ {\it SISSA--International School for Advanced Studies} \\
{\it via Beirut 2--4, I--34013 Trieste, Italy} \\ \\
$^{2\,}${\it Department of Physics, University of Oxford} \\
{\it Keble Road, Oxford OX1 3RH, UK} \\
{\it [Address after March 1, 1994]} \\ \\
$^{3\,}${\it Dipartimento di Astronomia, Universit\`a di Padova} \\
{\it vicolo dell'Osservatorio 5, I--35122 Padova, Italy} \\ }
\date{}
\begin{document}
\maketitle
\vspace{1.5cm}
\bc
{\it Astrophysical Journal}, submitted
\ec
\newpage
\baselineskip=0.5truecm       % lunghezza apjlett.
\vspace{1cm}
\bc
\section*{Abstract}
\ec
We discuss the non--linear growth of the kurtosis of the smoothed peculiar
velocity field (along an arbitrary direction), in an Einstein--de Sitter
universe, induced by Gaussian primordial density fluctuations. Applying the
perturbative theory, we show that, for different cosmological models, a
departure from the original Gaussian distribution is gravitationally induced
only on small scales ($\simlt 20-30$ Mpc). Models with scale--free power
spectrum $P(k)\propto k^{\,n}$, with $-1< n \leq 1$, are also considered. When
the fluid particles move according to the Zel'dovich approximation the
probability distribution of the peculiar velocity field remains unaltered
during the evolution.

\vspace{0.5cm}
\noindent{\em Subject headings:} Cosmology -- Galaxies: clustering --
large--scale structure of the Universe

\vspace{1.0cm}
\section{Introduction}
In the past decade astronomers and cosmologists devoted a great effort to
measure large--scale deviations from the Hubble flow and to interpret the
cosmological inferences. Redshift--independent distance estimators and all--sky
catalogs of galaxy redshifts allowed to estimate the (radial) peculiar motions
of galaxies. Undertaking detailed maps of the peculiar velocity field allows
for example to make dynamical estimates of the density parameter $\Om_{\circ}$,
to test the primordial density fluctuation power spectrum $P(k)$, and to
measure directly the underlying total (luminous plus dark) mass distribution,
once the gravitational instability picture is assumed.

The $statistical$ properties of the large--scale motions may be described in
terms of the peculiar velocity correlation tensor (G\'orski \etal 1989; Groth,
Juszkievicz, \& Ostriker 1989; Tormen \etal 1993) and/or the central moments of
the velocity probability distribution function (pdf) $p(\bfv)$ (Kofman \etal
1994; Bernardeau 1994). Assuming that the very early density pdf is Gaussian,
it results that, during the linear regime, $p(\bfv)$ is a Gaussian too, with
mean $\lan\bfv\ran={\bf 0}$ and variance
$\s_{\bfv}^{\,2}=(H_{\circ}^{\,2}\Om_{\circ}^{\,1.2}/2\pi^2)
\int_0^{\infty}dk\,P(k)\,$, $H_{\circ}$ being the Hubble constant. Even if the
primordial density field is Gaussian distributed, the non--linear time
evolution will ensure that the mass density fluctuations $\de$ become highly
non--Gaussian (Peebles 1980; Fry 1984; Goroff \etal 1986; Juszkiewicz, Bouchet,
\& Colombi 1993; Catelan \& Moscardini 1994), implying a modification of the
original Gaussian pdf $p(\bfv)$. However, due to the isotropy of the
cosmological velocity field, it is expected that $all\,$ the $odd\,$ moments of
$p(\bfv)$ remain zero during the growth of the density fluctuations. A
particular case is the third central moment, the velocity skewness, discussed
for example in Ruamsuwan \& Fry (1992). Thus, any gravitationally induced
departure from the Gaussian distribution may be sought only in the fourth
central moment of $p(\bfv)$, the velocity kurtosis, as well as in higher order
$even$ central moments (Grinstein \etal 1987; Kofman \etal 1994; Bernardeau
1994). These have been also analyzed in the framework of the global texture
model by Scherrer (1992) and Catelan \& Scherrer (1994).

In this work, we study the non--Gaussian content of the velocity pdf in terms
of the velocity kurtosis as induced by the gravitational growth of the
initially Gaussian density fluctuations. The velocity kurtosis describes
important features such as sharpness of the velocity pdf and the extent of its
rare--event tail. Taking advantage of the exact perturbative technique (Fry
1984; Goroff \etal 1986) and the Zel'dovich approximation (see Grinstein \&
Wise 1987), we estimate the kurtosis of the peculiar velocity along a direction
$\widehat{\al}$, namely the parameter $K_{v}\equiv [\lan
v_{\al}^{\,4}\ran-3\,\lan v_{\al}^{\,2}\ran^2]/ \lan v_{\al}^{\,2}\ran^2\,$,
after smoothing with a Gaussian filter. In particular we study the dependence
on both the scale in the context of several cosmological scenarios and the
primordial spectral index, when an initial (scale--free) power spectrum
$P(k)\propto k^{\,n}$ is assumed.

The layout of this paper is the following. In Section 2 the exact perturbative
theory and the (Eulerian version of the) Zel'dovich approximation are reviewed.
In Section 3 we discuss the gravitationally induced velocity kurtosis parameter
$K_{v}$ of an initially Gaussian peculiar velocity field. Our results and
conclusions are presented in Section 4.
\section{Non--Linear Time Evolution}
We assume that present--day structures formed by gravitational instability from
Gaussian fluctuations $\de$ in a pressureless fluid with matter density
$\rho=\rho_b[1+\de]$, where $\rho_b$ is the background mean density. The
density fluctuation field $\de$ may be written as a Fourier integral,
$
\de(\bfx,t)=(2\pi)^{-3}\int d\bfk\,\fd(\bfk,t)\,
{\rm e}^{\,i\,\bfk\cdot\bfx}\,,
$
where $\bfx$ and $\bfk$ are the {\it comoving} Eulerian coordinate and
wavevector, $t$ is the cosmic time. The power spectrum $P(k)$ fully determines
the statistics of the primordial Gaussian density field, whose variance is
$\s^{\,2}=(1/2\pi^2)\int_0^{\infty}dk\,k^2\,P(k)$. We filter the field $\de$ by
means of a Gaussian window function $W_R(x)=(2\pi
R^2)^{-3/2}\exp(-x^2/2R^2)\,$. The mass variance on scale $R$, $\s_R^{\,2}\,$,
is related to $P(k)$ by
$\s_R^{\,2}=(1/2\pi^2)\int_0^{\infty}dk\,k^2\,P(k)\,[\widetilde{W}_R(k)]^2\,$,
where $\widetilde{W}_R(k)$ is the Fourier transform of $W_R(x)\,$.
\subsection{Equations of Motion: Perturbative Theory}
The time evolution equations for the matter density fluctuation $\de(\bfx,t)$
and the peculiar velocity field $\bfv(\bfx, t)$ are the Euler equation and the
continuity equation, i.e.
\be
\p_{\circ}\bfv+\f{1}{a}\,(\bfv\cdot\nabla)\,\bfv+
\f{\dot{a}}{a}\,\bfv=\bfg \;,
\ee
\be
\p_{\circ}\de+\f{1}{a}\,\nabla\cdot(1+\de)\,\bfv=0\;.
\ee
Here $\p_{\circ}\equiv\p/\p t$ and spatial derivatives are with respect to
$\bfx$. The density contrast $\de$ is related to the Newtonian gravitational
potential
$
\tri(\bfx,t)\equiv-\,(4\pi)^{-1} \int d\bfx'\,\de(\bfx',t)/|\bfx'-\bfx|
$
via the Poisson equation, $\nabla^2\!\tri=\de\,$. In terms of $\tri$ the
peculiar gravitational acceleration is defined as $\bfg=-\,4\pi
G\,\rho_b\,a\,\nabla\!\tri\,$. We analyze these equations assuming an
Einstein--de Sitter universe with no cosmological constant. In such a model,
the scale factor $a$ is proportional to $t^{\,2/3}$ during the matter dominated
epoch, and the adiabatic expansion implies that $6\pi G\,\rho_b\,t^2 =1\,$.

The first--order  solution for $\de$ has the well--known self--similar form,
namely, considering only the growing mode,
$
\done(\bfx,t)=D(t)\,\de_1(\bfx)\,,
$
where $D(t)\propto a(t)$ is the time growth factor of the mass fluctuations. In
the linear regime, the peculiar velocity field is proportional to the
gravitational acceleration
\be
\bfv=-\,a\,\f{\dot{D}}{D}\,\nabla\!\tri\;.
\ee
This relation shows that the linear velocity field is irrotational; its growing
mode corresponds to the growing mode of the density field and, for a flat
universe, the classical law $\bfv=\bfg\,t\sim t^{\,1/3}\,$ is recovered.
According to Eq.(3), the particles of the gravitating fluid move along the
direction of the gravitational force.

It may be useful to give explicitly the Fourier transform of Eq.(3), i.e., for
the component along a fixed direction $\widehat{\al}$,
\be
\fv_{\al}(\bfk,t)=i\,a\,\f{\dot{D}}{D}\;\f{k^{\al}}{k^{\,2}}\;\fd(\bfk,t)\;.
\ee

Higher order approximations of the density solution may be recovered if one
expands the mass density fluctuation field $\de(\bfx, t)$ about the background
solution $\de=0\,$, namely $\de=\sum_n\de^{(n)}$ with
$\de^{(n)}=O(\de_1^{\,n})$, then solving the differential equation for any
$\de^{(n)}$ (Peebles 1980; Fry 1984). The perturbative expansion for $\de$
reads (e.g. Goroff \etal 1986):
$
\de(\bfx,t)=\sum_{n=1}^{\infty}[D(t)]^n\,\de_n(\bfx)\,.
$
The first term of the expansion corresponds to the linear approximation. We see
that the scale factor $D(t)$ acts as a coupling constant, since $\de^{(n)}
\propto D^n$.

In a similar fashion, expanding $\bfv$ about the solution $\bfv={\bf 0}$, one
obtains
\be
\bfv(\bfx,t)=\f{2}{3}\,\f{a}{t}\sum_{n=1}^{\infty}\,[D(t)]^n\,\bfv_n(\bfx)\;,
\ee
where $\bfv_n=O(\bfv_1^{\,n})\,$ and $\,-\,\nabla\cdot\bfv_1=\de_1\,$. We
assume that at any order $\nabla\wedge\bfv_n={\bf 0}\,$ (Kelvin circulation
theorem).

Here we review the exact perturbative technique to solve approximately the
equations of motion (1) and (2) explicitly up to third order in the peculiar
velocity field. In particular, we use the third--order solution to compute the
fourth--order moment (namely the kurtosis $K_{v}$). We adopt the same notation
of Fry (1984) and Catelan \& Moscardini (1994).
\bc
{\it (i) Second--Order Velocity Solution}
\ec
The second--order peculiar velocity $\bfv^{(2)}$ is a solution of the equations
\be
\p_{\circ}\left(a\,\bfv^{(2)}\right)+
\left(\bfv^{(1)}\cdot\nabla\right)\bfv^{(1)}=a\,\bfg^{(2)}\;,
\ee
\be
\p_{\circ}\,\de^{(2)}+a^{-1}\,\nabla\cdot
\left(\bfv^{(2)}+\de^{(1)}\,\bfv^{(1)}\right)=0\;,
\ee
where the second--order peculiar acceleration is $\bfg^{(2)}=-\,4\pi
G\,\rho_b\,a\,\nabla\!\tri^{(2)}\,$ and $\nabla^2\!\tri^{(2)}\equiv\de^{(2)}$.
The second--order density contribution $\de^{(2)}$ has been derived by Peebles
(1980). Since $\de^{(2)}\propto D^2\,$, it results that
$\bfg^{(2)}\propto\rho_b\,a\,D^2$ and the second--order velocity solution reads
\be
\bfv^{(2)}=-\,a\,\f{\dot{D}}{D}\,
\left[\,2\,\nabla\!\tri^{(2)}-\de^{(1)}\,\nabla\!\tri^{(1)}\,\right]
+\bfF_2\;,
\ee
where $\bfF_2$ is a divergenceless vector such that
$\nabla\wedge\bfv^{(2)}={\bf 0}$ (Catelan \etal 1994). The Fourier transformed
$\fv^{(2)}_{\al}$ may be directly obtained from the solution (8),
\be
\fv^{(2)}_{\al}(\bfk,t) =i\,a\,\f{\dot{D}}{D}\,\f{k^{\,\al}}{k^{\,2}}
\int \f{d\bfk'}{(2\pi)^3}\,
\,K^{(2)}(\bfk',\bfk-\bfk')\,\,\fd^{(1)}(\bfk',t)
\,\fd^{(1)}(\bfk - \bfk',t)\;,
\ee
where we have defined the kernel
\be
K^{(2)}(\bfk_1,\bfk_2) \equiv
\f{3}{7} + \f{\bfk_1\cdot\bfk_2}{k_2^{\,2}} +
\f{4}{7}\left(\f{\bfk_1\cdot\bfk_2}{k_1\,k_2} \right)^2\;.
\ee

It is straightforward to show that $\lan\bfv^{(2)}\ran={\bf 0}\,$. In an
Einstein--de Sitter universe $\bfv^{(2)}\sim t$, and it grows slower than
$\de^{(2)}\sim t^{\,4/3}$. We stress the fact that $\bfv^{(2)}$ is {\it not}
parallel to the second--order acceleration [$\propto -\,\nabla\!\tri^{(2)}\,$]:
this is a consequence of non--locality. Thus, the gravitational field changes
direction and the particles are not accelerated in a fixed direction, unlike
the linear regime. The density--velocity relation in the quasi--linear regime
and the cosmological implications of non--locality have been recently explored
by Nusser \etal (1991) and Gramann (1993a). The solution (9) is also derived by
Gramann (1993b), who applies a second--order Lagrangian perturbative technique.
\bc
{\it (ii) Third--Order Velocity Solution}
\ec
The third--order approximation $\bfv^{(3)}$ is a solution of the differential
equations
\be
\p_{\circ}\left(a\,\bfv^{(3)}\right)+
\left(\bfv^{(1)}\cdot\nabla\right)\bfv^{(2)}+
\left(\bfv^{(2)}\cdot\nabla\right)\bfv^{(1)}=a\,\bfg^{(3)}\;,
\ee
\be
\p_{\circ}\,\de^{(3)}+a^{-1}\,\nabla\cdot
\left(\bfv^{(3)}+\de^{(1)}\,\bfv^{(2)}+\de^{(2)}\,\bfv^{(1)}\right)
=0\;.
\ee
Here $\,\bfg^{(3)}=-\,4\pi G\,\rho_b\,a\,\nabla\!\tri^{(3)}\,$ and
$\,\nabla^2\!\tri^{(3)}\equiv\de^{(3)}$. The third--order density solution has
been obtained by Fry (1984). Since $\bfg^{(3)}\propto\rho_b\,a\,D^3$, and using
the results of the previous subsection,  the velocity $\bfv^{(3)}$ may be
written as
\be
\bfv^{(3)}=-\,a\,\f{\dot{D}}{D}
\left[\,
3\,\nabla\!\tri^{(3)}+
\de^{(1)\,2}\,\nabla\!\tri^{(1)}-
2\,\de^{(1)}\,\nabla\!\tri^{(2)}-
\de^{(2)}\,\nabla\de^{(1)}\,
\right]+\bfF_3\;.
\ee
Again the additive term $\bfF_3$ is such that $\nabla\wedge\bfv^{(3)}={\bf 0}$.
The Fourier transform of the previous expression is
$$
\widetilde{\bfv}^{(3)}(\bfk, t)=
$$
\be
i\,a\,\f{\dot{D}}{D}\,\f{\bfk}{k^{\,2}}
\int \f{d\bfk_1\,d\bfk_2\,d\bfk_3}{(2\pi)^6}
\,\de_D\Bigl(\,\sum_{h=1}^3\bfk_h - \bfk\Bigr)
\,\,K^{(3)}(\bfk_1, \bfk_2, \bfk_3)
\,\,\fd^{(1)}(\bfk_1,t)\,\fd^{(1)}(\bfk_2,t)
\,\fd^{(1)}(\bfk_3,t)\;,
\ee
where the third--order kernel is
\be
K^{(3)}(\bfk_1, \bfk_2, \bfk_3)=
3\,J^{(3)}(\bfk_1, \bfk_2, \bfk_3)-
\f{\bfk\cdot\bfk_1}{k_1^{\,2}}\,J^{(2)}(\bfk_2, \bfk_3)-
\f{\bfk\cdot(\bfk_1+\bfk_2)}{(\bfk_1+\bfk_2)^2}\,K^{(2)}(\bfk_1, \bfk_2)\;,
\ee
and the functions $J^{(2)}$ and $J^{(3)}$, corresponding to the second-- and
third--order density solutions (see e.g. Fry 1984; Catelan \& Moscardini 1994),
read respectively
\be
J^{(2)}(\bfk_1,\bfk_2) \equiv
\f{5}{7} + \f{\bfk_1\cdot\bfk_2}{k_2^{\,2}} +
\f{2}{7}\left(\f{\bfk_1\cdot\bfk_2}{k_1\,k_2} \right)^2\;,
\ee
$$
J^{(3)}(\bfk_1,\bfk_2,\bfk_3)\equiv J^{(2)}(\bfk_2, \bfk_3)
\left[\,\f{1}{3}+
\f{1}{3}\,\f{\bfk_1\cdot(\bfk_2+\bfk_3)}{(\bfk_2+\bfk_3)^2} +
\f{4}{9}\, \f{\bfk\cdot\bfk_1}{k_1^{\,2}} \,
\f{\bfk\cdot(\bfk_2+\bfk_3)}{(\bfk_2+\bfk_3)^2}
\,\right]
$$
\be
- \f{2}{9}\, \f{\bfk\cdot\bfk_1}{k_1^{\,2}}
\,\f{\bfk\cdot(\bfk_2+\bfk_3)}{(\bfk_2+\bfk_3)^2}
\,\f{(\bfk_2+\bfk_3)\cdot\bfk_3}{k_3^{\,2}}+
\f{1}{9}\, \f{\bfk\cdot\bfk_2}{k_2^{\,2}} \,\f{\bfk\cdot\bfk_3}{k_3^{\,2}}
\;.
\ee
It is not difficult to show that $\lan\bfv^{(3)}\ran={\bf 0}$. In an
Einstein--de Sitter universe, $\bfv^{(3)}\sim t^{5/3}\,$.
\bc
{\it (iii) General Solution}
\ec
As Goroff \etal (1986) have shown, the (Fourier transformed) $n$--th order
velocity solution may be represented in integral form as
\be
\widetilde{\bfv}_n(\bfk)=
i\,\f{\bfk}{k^{\,2}}
\left\{\prod_{h=1}^n\int\f{d\bfk_h}{(2\pi)^3}\,\fd_1(\bfk_h)\,
\right\}
\,\Bigl[(2\pi)^3\de_D(\,\sum_{j=1}^n\bfk_j-\bfk)\Bigr]\,K^{(n)}
(\bfk_1,\dots,\bfk_n)\;.
\ee
The presence of the Dirac delta function comes from momentum conservation in
Fourier space. The kernels $K^{(n)}$ are homogeneous (with degree 0) functions
of the wavevectors $\bfk_1,\ldots,\bfk_n\,$, describing the effects of
non--linear collapse (tidal and shear effects). In general the $K^{(n)}$ are
very
complicated for $n>3$. [A discussion of the properties of the kernels $K^{(n)}$
is given in Wise (1988). Explicit recursion relations with their Feynman
diagrammatic representations are given by Goroff \etal (1986) and Wise (1988).]
\subsection{Zel'dovich Approximation}
In the {\it Zel'dovich approximation} (Zel'dovich 1970) the motion of particles
from the initial comoving (Lagrangian) positions $\bfq\,$ is approximated by
straight paths. The Eulerian position at time $t$ is then given by the uniform
motion
\be
\bfx(\bfq,t)=\bfq+D(t)\,{\bf S}(\bfq)\;,
\ee
where $D(t)$ is the growth factor of linear density perturbations and ${\bf
S}(\bfq)$ is the displacement vector related to the primordial velocity field.
Grinstein \& Wise (1987) give an Eulerian representation of the Zel'dovich
approximation by a diagrammatic perturbative approach similar to that of the
previous section. They showed that the $n$--th order perturbative corrections
$\de_n(\bfx)$, when the density fluctuation field $\de$ is evolved according to
the Zel'dovich approximation, are such that
\be
\de(\bfx,t)=\sum_{n=1}^{\infty}\f{(-1)^n}{n!}\,[D(t)]^n
\sum_{[h_n]=1}^3\f{\p}{\p x_{h_1}}\cdots\f{\p}{\p x_{h_n}}\,
\Bigl[S_{h_1}\cdots S_{h_n}\Bigr]\;.
\ee
Here $\sum_{[h_n]}\equiv\sum_{h_1}\cdots\sum_{h_n}$. Note that the first term
recovers the linear approximation, in that ${\bf S}=\bfv_1$, where
$\de_1(\bfx)=-\nabla\cdot\bfv_1\,$. This expansion for $\de$ corresponds to
different symmetric kernels $K_{ZA}^{(n)}\,$, which can be written in the
following compact form
\be
K_{ZA}^{(n)}(\bfk_1,\ldots,\bfk_n)=
\f{1}{n!}\,\prod_{h=1}^n\,\f{\bfk\cdot\bfk_h}{k_h^{\,2}}\;,
\ee
where $\bfk\equiv\sum_{h=1}^n\bfk_h\,$. The kernels $\,K_{ZA}^{(n)}\,$ are the
same obtained in the expansion of the density contrast (see Catelan \&
Moscardini 1994) and are symmetric by construction.
\section{\bf Kurtosis of the Velocity Field}
In this section, we compute the gravitationally induced kurtosis $K_{v}$ of an
initial Gaussian velocity field in a flat universe. We restrict the calculation
to the velocity along a chosen direction $\widehat{\al}\,$. The lowest order
non--zero reduced contribution to $K_{v}$ is
\be
\lan v_{\al}^{(1)\,2}\ran^2\,K_{v}\,\equiv\,6\,\lan
\,v_{\al}^{(1)\,2}\,v_{\al}^{(2)\,2}\,\ran_c
\,+\,4\,\lan\,v_{\al}^{(1)\,3}\,v_{\al}^{(3)}\,\ran_c\;,
\ee
where the subscript $c$ indicates the connected part of the four--point
velocity correlation.

It is not difficult to verify that $K_{v}$ depends on the normalization of the
power spectrum. Furthermore it varies with time. The related quantity which is
independent both of the power spectrum normalization and of time is $S_{4,v}
\equiv K_v / \lan \de^{\,(1)2}\ran$: due to these properties, observational
estimates of $S_{4,v}$ would allow one to detect $\,intrinsic\,$ features of
the evolved large--scale velocity field.

{}From the perturbative results in Eqs.(9), (14) and (22), one finally gets the
integral expression of the kurtosis of the smoothed velocity field $\bfv_R$:
$$
K_{v}(R) = \f{24}{ \s_{v,R}^{\,4} \,}
\left(a\,\f{ \dot{D} }{ D }\right)^4
\int \f{d\bfk_1\,d\bfk_2\,d\bfk_3}{(2\pi)^9}\,\,
\f{k_1^{\al}\,\,k_2^{\al}\,\,k_3^{\al}\,\,(-\,k_1^{\al}-k_2^{\al}-k_3^{\al})}
{ k_1^{\,2}\,\,k_2^{\,2}\,\,k_3^{\,2}\,|\bfk_1+\bfk_2+\bfk_3|^{\,2} }
\,\times
$$
$$
\widetilde{W}_R(k_1)\,\,\widetilde{W}_R(k_2)\,\,\widetilde{W}_R(k_3)\,\,
\widetilde{W}_R(|\bfk_1+\bfk_2+\bfk_3|)\,\times
$$
\be
P(k_1)\,P(k_2)
\left[P(k_3)\,K_s^{(3)}(\bfk_1, \bfk_2, \bfk_3)+
2\,K_s^{(2)}(-\bfk_2, \bfk_2+\bfk_3)\,
K_s^{(2)}(\bfk_1, \bfk_2 + \bfk_3)\,P(|\bfk_2+\bfk_3|)\right]\,,
\ee
where $\s^2_{v,R}$ is the variance of smoothed velocity field. The kernels
$K_s^{(n)}$ are obtained by complete symmetrization of the kernels $K^{(n)}$.
The analogous expression in the Zel'dovich approximation is obtained by
replacing the kernels $K_s^{(n)}$ with the corresponding $\,K^{(n)}_{ZA}\,$.
Finally, note that $P(k)$ completely describes the process of growth of the
higher order velocity moments from Gaussian initial conditions.
\section{Discussion and Conclusions}
We calculate the previous integrals, by an Adaptive Multidimensional Monte
Carlo Integration subroutine, in the framework of several different
cosmological models. In particular, we consider: (1) standard cold dark matter
model (SCDM), i.e. with $b=1$ for the linear biasing parameter; (2) a biased
(BCDM) version of the same model, with $b=1.5$; (3) tilted cold dark matter
model (TCDM), with spectral index $n=0.7$ and $b=2.0$; (4) a `mixed' model
(MDM), with 60\% cold and 30\% hot dark matter, $b=1.5$. All transfer functions
have been taken from Holtzmann (1989), with the Hubble constant $H_\circ=50$ km
s$^{-1}$ Mpc$^{-1}$.

The results for $K_v$ are shown in Fig.1a. The associate relative uncertainty
estimates from the Monte Carlo Integration (not shown in the figure for
clarity) are always smaller than 5\%. For any model the velocity kurtosis is a
decreasing function of the smoothing scale $R$. On large scales $K_v$ is always
consistent with zero, thus the distribution of the peculiar velocity field is
very close to a Gaussian one. However, a strong non--Gaussian signature -- i.e.
$K_v \simgt 1$ -- is induced on scales smaller than $\simlt 20-30$ Mpc, where
nevertheless observations are at the moment affected by large uncertainties.
Furthermore, even if the general trend is similar for all considered models,
the departure from Gaussianity is larger for models with low biasing parameter:
TCDM gives always the lowest values of $K_v$.

A previous estimate of the kurtosis of the velocity field was obtained by
Kofman \etal (1994), who used the smoothed velocity fields of a $N$--body
simulation of the standard cold dark matter model with box--size of 400 Mpc to
study the velocity distribution $p(\bfv)$. In order to have a more direct
comparison, in this case we calculated $K_v$ using their choice for the CDM
transfer function (Davis \etal 1985) and limiting the numerical integration of
the expression (23) in the range of wavevectors spanned by the $N$--body
simulation: from the frequency corresponding to the box--size to the Nyquist
one. In Fig.1b we show both our perturbative evaluations and the Kofman et
al.'s data: the agreement is quite good, except for the small scale (12 Mpc)
result. It might seem surprising that the numerical simulations, which are
supposed to better describe the highly nonlinear regime, miss the departure
from Gaussianity of the velocity field: in fact, for the density field, the
perturbative evaluations of higher moments underestimate the $N$--body results.
A possible explanation is that the multiple filtering (a trilinear
interpolation plus a small--scale smoothing) necessary in the simulations to
reconstruct the velocity field from the particle distribution might smooth the
small--scale non--Gaussian signal. Moreover, a further ``numerical smoothing"
can appear due to the intrinsic gridding present in the particle--mesh code. In
any case, it is necessary to be cautious in the measurements of higher order
moments in $N$--body simulations because they are strongly affected by the
high--value tails, which characterize the particular realization, size of the
box and other numerical problems (Kofman, private communication): this is also
shown by the large error bars in the $N$--body results.

We also calculate the velocity kurtosis in terms of the intrinsic parameter
$S_{4,v}$. In Fig.2a, $S_{4,v}$ is plotted for the same models previously
considered: in this case, due to the independence of $S_{4,v}$ on the power
spectrum normalization, the biased and the standard CDM originate the same
curve. All models present a similar behavior and the differences are always
inside the error bars, shown for clarity only for SCDM. In Fig.2b, the
dependence of $S_{4,v}$ on the primordial spectral index $n$ is shown for
scale--free power spectra $P(k) \propto k^{\,n}$, with $n$ in the range $-1 < n
\le 1$, for both the perturbative and Zel'dovich approximations. Due to the
assumed scale--invariance, $S_{4,v}$ only depends on the primordial spectral
index $n$, and not on the scale $R$. It may be noted that in the Zel'dovich
approximation, unlike the perturbative case, the velocity kurtosis is
practically constant and consistent with zero for any value of $n$. Thus, we
confirm the results of Kofman \etal (1994), who demonstrate that the Eulerian
Gaussian one--point pdf $p(\bfv)$ is {\it time--invariant} as long as the
Zel'dovich approximation holds. This is essentially due to the simple time
scaling of the particle Eulerian position $\bfx(\bfq,t)$ in Eq.(19).

Finally we want to stress that one has to be careful about making quantitative
comparisons between our results, directly related to the underlying (dark plus
luminous) mass distribution, and observational data. In particular, Grinstein
\etal (1987) suggest that when point--like luminous objects, like galaxies, are
used to sample the peculiar flow within a region of the sky, it is the volume
average of $n(\bfx)\bfv(\bfx)$ which is actually measured instead of the volume
average of $\bfv(\bfx)$, $n(\bfx)$ being the number density of luminous
tracers. If the objects are biased tracers of the underlying mass distribution,
then nonlinear effects on small--scales may preclude any direct comparison of
the observed velocity moments with ensemble expectations. A further analysis of
this problem is in progress.

\vspace{1.cm}
\noindent
{\normalsize \bf Acknowledgments\,}
We are very grateful to Lev Kofman, Sabino Matarrese and Mark Wise for many
fruitful discussions. PC acknowledges Robert Scherrer for his kind hospitality
at the Ohio State University, Columbus, where part of this work begun. This
work has been supported by Italian MURST.

\vspace{1.cm}
\newpage
\parindent 0pt
\section*{References}

\begin{trivlist}

\item [] Bernardeau, F. 1994, ApJ, in press

\item[] Catelan, P., Lucchin, F., Matarrese, S., \& Moscardini, L. 1994,
in preparation

\item[] Catelan, P., \& Moscardini, L. 1994, ApJ, in press

\item[] Catelan, P., \& Scherrer, R.J. 1994, in preparation

\item[] Davis, M., Efstathiou, G., Frenk, C., \&  White, S.D.M.
1985, ApJ, 292, 371

\item[] Fry, J.N. 1984, ApJ, 279, 499

\item[] Goroff, M.H., Grinstein, B., Rey, S.--J., \& Wise, M.B. 1986, ApJ,
	311, 6

\item[] Gorski, K., Davis, M., Strauss, M.A., White, S.D.M., \& Yahil,
A. 1989, ApJ, 344, 1

\item[] Gramann, M. 1993a, ApJ, 405, 449

\item[] Gramann, M. 1993b, ApJ, 405, L47

\item[] Grinstein, B., Politzer, H.D., Rey, S.--J., \& Wise, M.B. 1987,
	ApJ, 314, 431

\item[] Grinstein, B., \& Wise, M.B. 1987, ApJ, 320, 448

\item[] Groth, E.J., Juszkiewicz, R., \& Ostriker, J.P. 1989, ApJ, 346, 558

\item[] Holtzman, J.A. 1989,  ApJS, 71, 1

\item[] Juszkiewicz, R., Bouchet, F.R., \& Colombi, S. 1993, ApJ, 412, L9

\item[] Kofman, L., Bertschinger, E., Gelb, J.M., Nusser, A., \&
        Dekel, A. 1994, ApJ, 420, 44

\item[] Nusser, A., Dekel, A., Bertschinger, E., \& Blumenthal, G.R. 1991,
        ApJ, 379, 6

\item[] Peebles, P.J.E. 1980, {\it The Large Scale Structure of the Universe}
        (Princeton: Princeton Univ. Press)

\item[] Ruamsuwan, L., \& Fry, J.N. 1992, ApJ, 396, 416

\item[] Scherrer, R.J. 1992, ApJ, 390, 330

\item[] Tormen, G., Moscardini, L., Lucchin, F., \& Matarrese, S. 1993,
        ApJ, 411, 16

\item[] Wise, M.B. 1988, in {\it The Early Universe}, eds W.G. Unruh \&
        G.W. Semenoff, Reidel Publishing Company

\item[] Zel'dovich, Ya.B. 1970, A\&A, 5, 160

\end{trivlist}

\newpage

\section*{\center Figure captions}

{\bf Figure 1.} The kurtosis ratio $K_v$ of the velocity field. Left panel: the
behavior versus the scale $R$ for Gaussian filter and different cosmological
models (parameters are in the text): standard cold dark matter (solid line);
biased cold dark matter (dotted line); tilted cold dark matter (dashed line);
mixed dark matter (dotted--dashed line). Right panel: comparison with results
from $N$--body (Kofman \etal 1994, filled squares) in the case of standard cold
dark matter. Filled triangles and solid line refer to the perturbative
estimates when the integration is limited in the interval ranging from the
box--size frequency to the Nyquist one; dotted line refers to the perturbative
estimates when the integration is over all frequencies.

\bigskip

{\bf Figure 2.} The intrinsic parameter $S_{4,v}$. Left panel: the behavior
versus the scale $R$ for different cosmological models: standard cold dark
matter (triangles), tilted cold dark matter (squares), mixed dark matter
(circles). Error bars, shown only for SCDM, refer to the associated uncertainty
estimate from the Monte Carlo Integration. Right panel: the behavior versus the
primordial spectral index $n$ for power--law spectra $P(k)\propto k^{\,n}$ and
Gaussian filter, for both the perturbative (triangles and solid line) and
Zel'dovich approximations (squares and dotted line).

\end{document}